\documentclass[structabstract]{aa}
\usepackage{graphicx}
\usepackage{txfonts}
\begin{document}
\renewcommand{\textfraction}{0.0}
\renewcommand{\bottomfraction}{1.0}
\renewcommand{\dblfloatpagefraction}{1.00}

   \title{Triggered massive and clustered star formation by combined H {\Large II} regions G38.91-0.44 and G39.30-1.04}
  \author{Jin-Long Xu,
          \inst{1,2}
          Jun-Jie Wang\inst{1,2}
          \ and Xiao-Lan Liu\inst{1,2,3}
          }
   \institute{ National Astronomical Observatories, Chinese Academy of Sciences,
             Beijing 100012, China \\
         \email{xujl@bao.ac.cn}
         \and
          NAOC-TU Joint Center for Astrophysics, Lhasa 850000, China\\
        \and
            University of the Chinese Academy of Sciences, Beijing, 100080,  China\\
        }
\authorrunning{J.-L. Xu et al.}
\titlerunning{Triggered massive and clustered stars formation by together H {\footnotesize II} regions G38.91-0.44 and G39.30-1.04}
   \abstract
   {}
{We investigate the triggered star formation occurring in the Infrared dark clouds (IRDC) G38.95-0.47 between H {\small II} regions G38.91-0.44 and G39.30-1.04, and study the detailed morphology, distribution, and
physical parameters of the molecular gas and dust in this region.}
{We present the radio continuum, infrared, and CO molecular observations of infrared dark cloud (IRDC) G38.95-0.47 and its adjacent H {\footnotesize II} regions G38.91-0.44 (N74), G38.93-0.39 (N75), and G39.30-1.04.  The Purple Mountain Observation (PMO) 13.7 m radio telescope was used to detect $^{12}$CO $J$=1-0, $^{13}$CO $J$=1-0 and C$^{18}$O $J$=1-0 lines. The carbon monoxide (CO) molecular observations can ensure the real association between the ionized gas and the neutral material observed nearby. To select young stellar objects (YSOs) associated this region,  we used the GLIMPSE I catalog.}
{ The $^{13}$CO $J$=1-0 emission presents two large cloud clumps. The clump consistent with IRDC G38.95-0.47 shows a triangle-like shape, and has a steep integrated-intensity gradient toward H {\footnotesize II} regions G38.91-0.44 and G39.30-1.04,  suggesting that the two H {\footnotesize II} regions have expanded into the IRDC. Four submillmeter continuum sources have been detected in the IRDC G38.95-0.47. Only the G038.95-00.47-M1 source with a mass of 117 $M_{ \odot}$  has outflow and infall motions, indicating a newly forming massive star. We detected a new collimated outflow in the clump compressed by G38.93-0.39. The derived ages of the three H {\small II} regions are 6.1$\times10^{5}$ yr, 2.5$\times10^{5}$ yr, and 9.0$\times10^{5}$ yr, respectively. In the IRDC G38.95-0.47, the significant enhancement of several Class I YSOs indicates the presence of some recently formed
stars. Comparing the ages of these H {\small II} regions with YSOs (Class I sources and massive G038.95-00.47-M1 source), we suggest that YSOs may be triggered by  G38.91-0.44 and G39.30-1.04 together, which supports the radiatively driven implosion model. It may be the first time that the triggered star formation has occurred in the IRDC compressed by two H {\small II} regions. The new detected outflow may be driven by a star cluster. }
   {}


   \maketitle
%
%________________________________________________________________

\section{Introduction}
Massive stars play an important role in both the morphology and
the chemical evolution of the interstellar medium (ISM), but their formation mechanism is still poorly understood. The formation of massive stars can be triggered by the action of H {\small II} regions. With the expansion of the H {\small II} regions, it may compress a pre-existing molecular cloud creating a compact clump, or sweep up the surrounding molecular material creating a dense molecular layer.
The compact clump and dense molecular layer can be break apart and  then may collapse to lead to the formation of new stars.

In the numerical study of star formation, the expanding H {\small II} regions can trigger star formation if the ambient molecular material is dense enough (Hosokawa \& Inutska \cite{Hosokawa05}; Dale et al. \cite{Dale07}). Collect and collapse (CC) and radiatively driven implosion (RDI) mechanisms may trigger star formation on the borders of H {\small II} regions (Elmegreen \cite{Elmegreen98}). Several pieces of observational evidence have been found supporting these star formation mechanisms (e.g., Zavagno et al. \cite{Zavagno07}; Deharveng et al. \cite{Deharveng08}; Paron et al. \cite{Paron09}; Tibbs et al. \cite{Tibbs12}).
Therefore, a detailed study of the
star-forming regions at various wavebands is necessary to
trace the star formation
scenario of  H {\small II} region environments.

Infrared dark clouds (IRDCs) are considered potential cluster and
massive star formation regions (Menten et al. \cite{Menten}).
In this paper, we present results
of an IRDC G38.95-0.47 with a distance of $\sim$2.9 kpc (Du \& Yang \cite{Du}), located between
H {\footnotesize II} regions G38.91-0.44 and G39.30-1.04. Because IR-dark clumps are absorption features against the mid-IR Galactic background emission, 	
L\'{o}pez-Sepulcre et al. (\cite{Lopez-Sepulcre10}) suggested that IRDC G38.95-0.47 is most likely located at the near distance.	
Four millimeter continuum cores were detected in the IRDC G38.95-0.47 (Rathborne et al. \cite{Rathborne}).
Only the G038.95-00.47-M1 core has outflow and infall motions (L\'{o}pez-Sepulcre et al. \cite{Lopez-Sepulcre10}).  However, Alexander et al. (\cite{Alexander}) did not detect the classical signposts of triggered star formation in this region and concluded that the triggered star formation has not occurred.
The region G39.30-1.04 may be a newly identified H {\footnotesize II} region, while H {\footnotesize II} regions G38.91-0.44 and G38.93-0.39 are associated with the infrared bubbles N74 and N75, respectively (Churchwell et al. \cite{Churchwell}; Deharveng et al. \cite{Deharveng10}; Sherman \cite{Sherman}).

To look for signatures of star formation, we  combined
molecular, infrared, and radio continuum
observations toward IRDC G38.95-0.47 and its adjacent H {\footnotesize II} regions.

\begin{figure*}[]
\vspace{0mm}
\includegraphics[angle=0,scale=0.65]{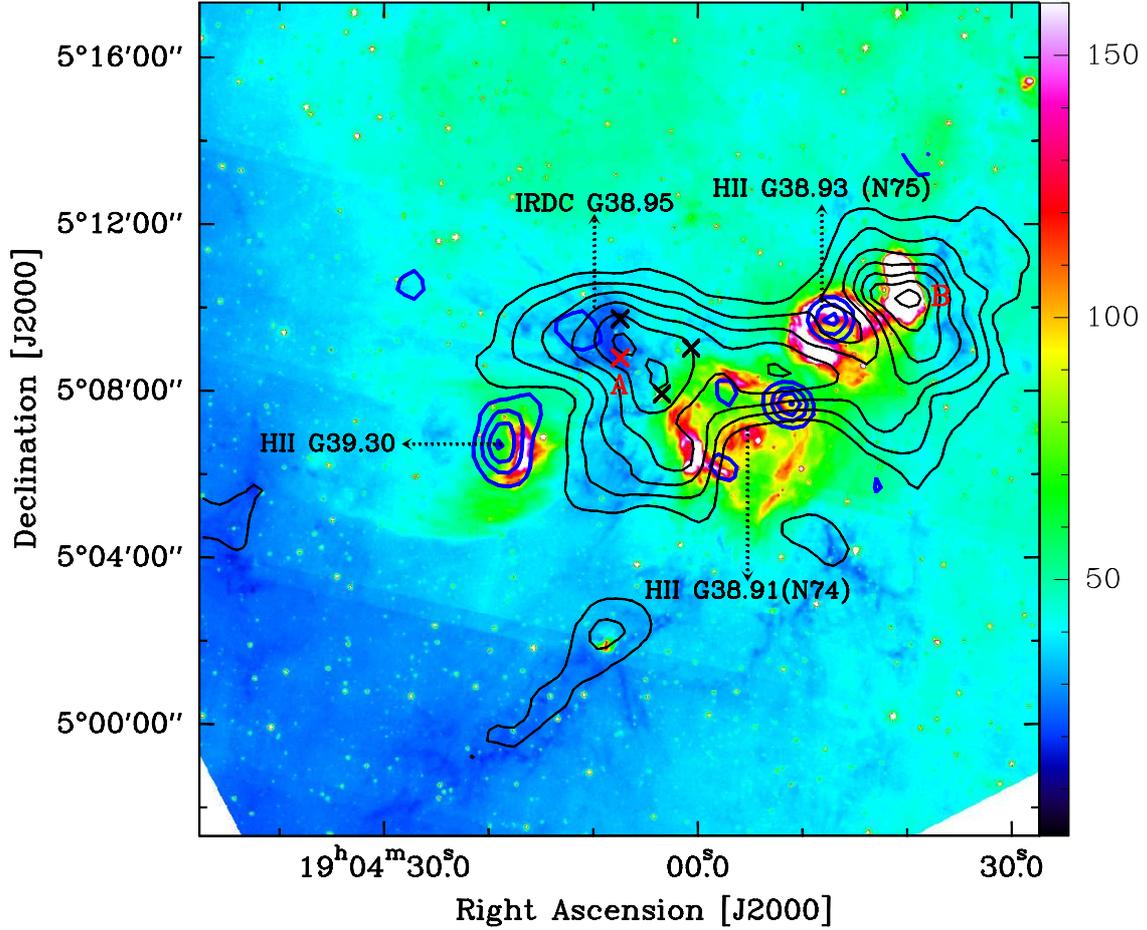}
\vspace{0mm}\caption{$^{13}$CO $J$=1-0 integrated intensity
contours (black) and 1.4 GHz radio continuum emission
contours (blue) overlayed on the Spitzer-IRAC 8
$\mu$m emission map (color scale). The black contour
levels are 30, 40,... , $90\%$ of the peak value (46.7 K km
s$^{-1}$). The blue contour levels are 1.7, 3.3, 4.9, 6.5, and 8.1
mJy beam$^{-1}$. The letters A and B
indicate the different cloud clumps. The green and red crosses indicate four  millimeter continuum sources.}
\end{figure*}

\begin{figure*}[]
\vspace{-24mm}
\includegraphics[angle=270,scale=1.08]{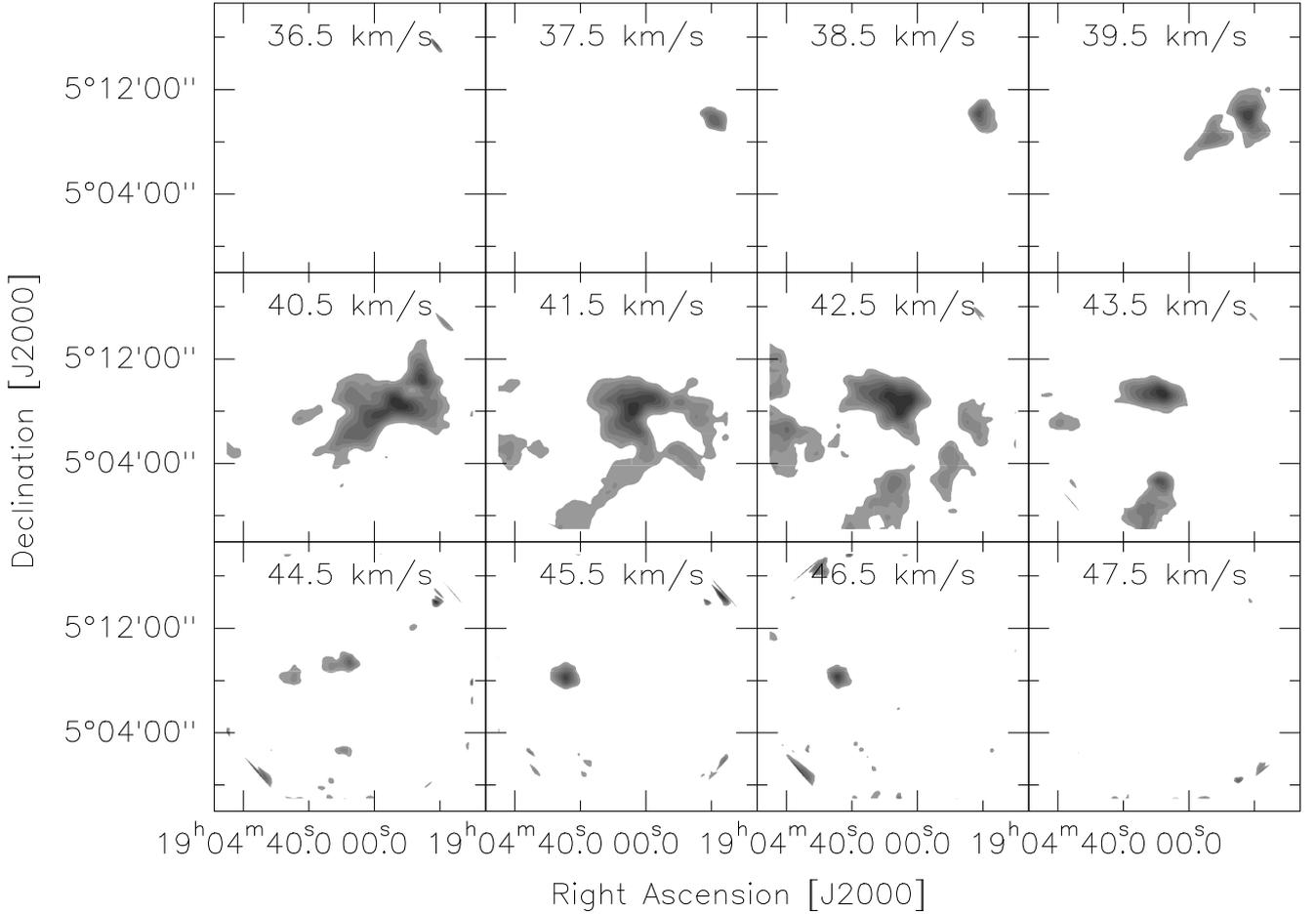}
\vspace{-42mm}\caption{$^{13}$CO $J$=1-0 channel maps each 1 km $s^{-1}$. Central velocities are indicated in
each image.}
\end{figure*}

\section{Observations and data reduction}
The mapping observations of IRDC G38.95-0.47 and its adjacent H {\footnotesize II} regions were performed in the $^{12}$CO
$J$=1-0, $^{13}$CO $J$=1-0, and C$^{18}$O $J$=1-0 lines using the Purple Mountain Observation (PMO) 13.7 m radio telescope at De Ling Ha, China, in May  2012.
The new 3$\times$3 beam array receiver system in single-sideband (SSB) mode was used as
front end. The back end is a fast Fourier transform spectrometer (FFTS) of 16384 channels with a bandwidth of 1 GHz,
corresponding to a velocity resolution of 0.16 km s$^{-1}$ for $^{12}$CO
$J$=1-0 and 0.17 km s$^{-1}$
for $^{13}$CO $J$=1-0 and C$^{18}$O $J$=1-0; $^{12}$CO
$J$=1-0 was observed at upper
sideband, while $^{13}$CO $J$=1-0 and C$^{18}$O $J$=1-0 were observed simultaneously
at lower sideband.  The half-power beam width (HPBW)
was 53$^{\prime\prime}$ at 115 GHz
 and the main beam efficiency was 0.5. The pointing
accuracy of the telescope was better than 5$^{\prime\prime}$. The system noise
temperature (Tsys) in SSB mode varied between 150 K and 400
K. Mapping observations were centered at
RA(J2000)=$19^{\rm h}04^{\rm m}07.5^{\rm s}$,
DEC(J2000)=$05^{\circ}08'18.9^{\prime\prime}$ using on-the-fly (OTF) observing mode. The total mapping area is $20^{\prime}\times 20^{\prime}$ in
$^{12}$CO $J$=1-0, $^{13}$CO $J$=1-0, and C$^{18}$O $J$=1-0 with a $0.5^{\prime}\times0.5^{\prime}$
grid. The standard chopper wheel calibration technique is used to measure antenna temperature $T_{\rm A}$$^{\ast}$ corrected for atmospheric absorption. The final data was recorded in brightness temperature scale of $T_{\rm mb}$ (K). The data were reduced using the GILDAS/CLASS\footnote{http://www.iram.fr/IRAMFR/GILDAS/} package. The 1.4 GHz radio continuum emission data were obtained from the
NRAO VLA Sky Survey  (Condon et al. \cite{Condon98}).

\section{Results}
\subsection{Radio continuum and infrared emission of H {\footnotesize II} regions}
Figure 1 shows the 1.4 GHz continuum emission image (blue contours) superimposed on the Spitzer-IRAC emission at 8 $\mu$m (color scale).
The Spitzer-IRAC 8 $\mu$m emission is attributed to polycyclic aromatic hydrocarbons (PAHs), which can be used to trace the photodissociated region (PDR) surrounding H {\footnotesize II} region.
In Fig. 1, the PAH emission displays a ring-like shape for the H {\footnotesize II} region G38.91-0.44, which coincides with bubble N74 (Churchwell et al. \cite{Churchwell}; Deharveng et al. \cite{Deharveng10}; Sherman \cite{Sherman}).
A small area of higher flux is situated at the center of the bubble, which could be a feature in the PDR on the far or near side of the bubble.
Moreover, one compact and two weak  radio continuum emissions (blue contours) are distributed along the PAH emission of G38.91-0.44.  Anderson et al. (\cite{Anderson11}) determined that the hydrogen radio recombination line (RRL) velocity is 40 km s$^{-1}$ for G38.91-0.44.  Hence, the triggered star formation is taking place in the molecular cloud at the periphery of PDR.
There may be another new H {\footnotesize II} region to the east of G38.91-0.44, named  G39.30-1.04 with RA(J2000)=$19^{\rm h}04^{\rm m}19.0^{\rm s}$ and
DEC(J2000)=$05^{\circ}06'42.7^{\prime\prime}$. The 1.4 GHz continuum emission reveals the presence of the ionized gas of G39.30-1.04. The PAH emission of G39.30-1.04 with a higher surface brightness shows a comet-like morphology. Because the ionized gas of G39.30-1.04 is surrounded by the PAH emission, the PAH emission may be excited by radiation from G39.30-1.04. The H {\footnotesize II} region G38.93-0.39 associated with bubble N75 (Churchwell et al. \cite{Churchwell}; Deharveng et al. \cite{Deharveng10}; Sherman \cite{Sherman}) is located at the northwest of G38.91-0.44. From Fig. 1, we can see that the PAH emission of G38.93-0.39 show an almost semi-ring shape with a cut towards the southwest.
The dense continuum emission is filled in the PAH emission of G38.93-0.39.

\subsection{Molecular line emission}
To analyze in greater detail the morphology of molecular gas associated with H {\footnotesize II} regions, we use the optically thin $^{13}$CO $J$=1-0 to trace the molecular gas.
After a careful inspection of the CO component using the channel map (see Fig. 2), we find that only the CO component in intervals 37 $\sim$ 47 km s$^{-1}$ is associated with these H {\footnotesize II} regions. Using this velocity range, we make the integrated intensity map of $^{13}$CO $J$=1-0 (Fig. 1).
From Fig. 1 (black contours), we find two large cloud clumps,  designated clumps A and B, respectively. Clump A covers the whole IRDC G38.95-0.47, which is located between H {\footnotesize II} region G38.91-0.44 and G39.30-1.04. In addition, four  submillimeter continuum sources were detected in the clump A (Di Francesco et al. \cite{Francesco08}). Only G038.95-00.47-M1 core  (marked by a red cross in figures 1 and 4) has outflow and infall motions (L\'{o}pez-Sepulcre et al. \cite{Lopez-Sepulcre10}). Moreover, $^{13}$CO $J$=1-0 emission of  clump A presents a triangle-like shape, and has an integrated intensity gradient along the direction of H {\footnotesize II} regions G38.91-0.44 and G39.30-1.04, suggesting that the shocks from the two H {\footnotesize II} regions have expanded into clump A, and have compressed it. Clump B shows a bow-like shape toward G38.93-0.39, the center of which has a small area of higher PAH emission.  Mercer et al. (\cite{Mercer}) detected an star cluster in clump B.

Figure 3 shows the average spectra of $^{12}$CO
$J$=1-0, $^{13}$CO $J$=1-0, and C$^{18}$O $J$=1-0 over clumps A and B, respectively. From these spectra, we can see
that the velocity components are mainly located
in the velocity interval 30 to 50 km s$^{-1}$. The velocity component in interval 10--20 km s$^{-1}$ should belong to the foreground emission, while 50--90 km s$^{-1}$ may belong to the background  emission, as seen in Fig. 2. We perform
Gaussian fits to the spectra of $^{12}$CO
$J$=1-0, $^{13}$CO $J$=1-0, and C$^{18}$O $J$=1-0 in clumps A and B. The fitted results are summarized
in Table 1. Hence, we derive systemic velocities of $\sim$41.7
km s$^{-1}$ and $\sim$39.4 km s$^{-1}$ in the C$^{18}$O $J$=1-0 lines for
clumps A and B, respectively, which are well associated with the hydrogen RRL velocity of H {\footnotesize II} region G38.91-0.44.

\begin{figure}[]
\vspace{-6mm}
\includegraphics[angle=270,scale=.68]{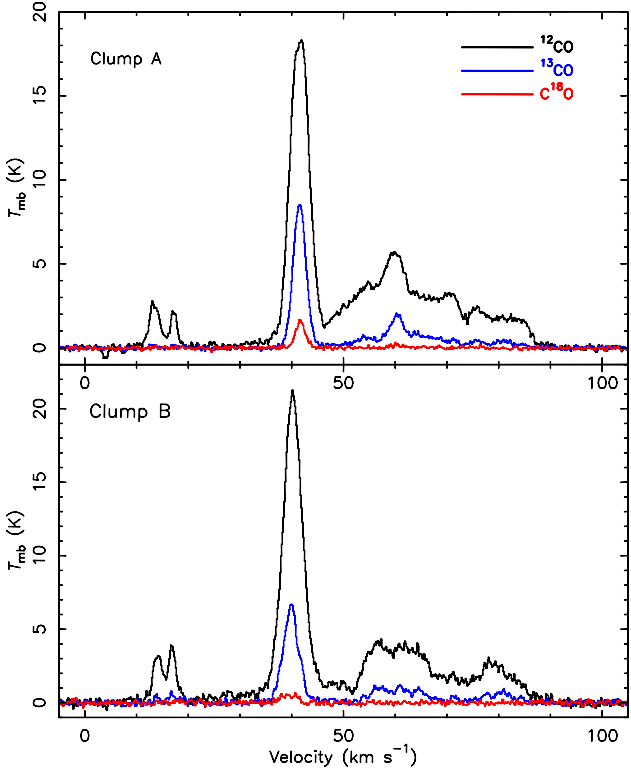}
\vspace{-16mm}\caption{Average spectra of $^{12}$CO
$J$=1-0, $^{13}$CO $J$=1-0, and C$^{18}$O $J$=1-0 over each clump.}
\end{figure}

\begin{table*}
\begin{center}
\tabcolsep 2.8mm\caption{Observed parameters of each line}
\def\temptablewidth{10\textwidth}%
\vspace{-2mm}
\begin{tabular}{lcccccccccc}
\hline\hline\noalign{\smallskip}
Name   &      & $^{12}$CO J=1-0   &   &   & $^{13}$CO J=1-0 &  & & C$^{18}$O J=1-0& \\
\cline{2-10}
        &   $T_{\rm mb}$   &FWHM  &$V_{\rm LSR}$   &   $T_{\rm mb}$   &FWHM  &$V_{\rm LSR}$&   $T_{\rm mb}$   &FWHM  &$V_{\rm LSR}$   \\
        &  (K)               &(km $\rm s^{-1}$) &(km $\rm s^{-1}$) &   (K)   &(km $\rm s^{-1}$)  &(km $\rm s^{-1}$)&   (K)   &(km $\rm s^{-1}$)&(km $\rm s^{-1}$) \\
\hline\noalign{\smallskip}
Clump A  & 18.8 & 4.4 (0.2)  & 41.6 (0.1)  &  8.6   & 2.9(0.1) & 41.5 (0.1)  & 1.6  & 2.2 (0.1) & 41.7 (0.1) \\  % new variable
Clump B  & 20.3 & 4.6 (0.2)  & 40.2 (0.1)  &  6.3   & 3.6(0.1) & 38.9 (0.1)  & 0.5  & 3.7 (0.2) & 39.4 (0.1) \\  % new variable
\noalign{\smallskip}\hline
\end{tabular}\end{center}
\end{table*}

 Assuming local thermodynamical equilibrium (LTE) and using the optically thin
$^{13}$CO $J$=1-0,  the column densities of the clumps
 are determined by the Garden et al. (\cite{Garden}) equation
\begin{equation}
\mathit{N_{\rm ^{13}CO}}=4.6\times10^{13}\frac{(T_{\rm
ex}+0.89)}{\exp(-5.29/T_{\rm ex})}\int T_{\rm mb}\rm dv ~\rm cm^{-2},
\end{equation}\indent
where $T_{\rm ex}$ is
the excitation temperature in K, and $\rm dv$ is the velocity range in km s$^{-1}$.  We calculate  $T_{\rm ex}$ following the
equation $T_{\rm ex}=5.53/{\ln[1+5.53/(T_{\rm mb}+0.82)]}$,
where $T_{\rm mb}$ is the corrected main-beam temperature of $^{12}$CO
$J$=1-0.  Here we
use the relation $N_{\rm H_{2}}$ $\approx$
$5\times10^{5}N_{\rm ^{13}CO}$  (Simon et al. \cite{Simon01}). If the clumps are
approximately spherical in shape, the mean number density of $\rm
H_{2}$ is estimated to be $n(\rm H_{2})$=$1.62\times10^{-19}N_{\rm H_{2}}/d$, where
$d$ is the averaged diameter of the clumps in parsecs (pc). Moreover, their mass
is given by $M_{\rm H_{2}}$=$\frac{1}{6}\pi
d^{3}\mu_{g}m(\rm H_{2})$$n(\rm H_{2})$ (Garden et al. \cite{Garden}), where
$\mu_{g}$=1.36 is the mean atomic weight of the gas, and $m(\rm
H_{2})$ is the mass of a hydrogen molecule. The obtained column
density, mean number density, and mass of each clump are all listed in Table 2.

\begin{table}[h!]
\vspace{-4mm}
\begin{center}
\tabcolsep 2.5mm \caption{The physical parameters of the clumps in LTE.}
\def\temptablewidth{0.5\textwidth}
\vspace{-2mm}
\begin{tabular}{ccccccccc}
\hline\hline\noalign{\smallskip}
Name   &$T_{\rm ex}$& $d$ &$N_{\rm H_{2}}$ & $n(\rm H_{2})$& $M$   \\
       & K&  pc  &(cm$^{-2}$) &(cm$^{-3}$)  & ($10^{5}\rm M_{ \odot}$)     \\
  \hline\noalign{\smallskip}
Clump A  &22.3 &  31.2 & 1.7$\times10^{22}$  & 0.9$\times10^{2}$ & 1.0   \\  % new variable
Clump B  &23.7 &  34.8 & 1.6$\times10^{22}$  & 0.8$\times10^{2}$ & 1.1  \\  % new variable
\noalign{\smallskip}\hline
\end{tabular}\end{center}
\vspace{-3mm}
\end{table}

\begin{figure*}[]
\vspace{-31mm}
\includegraphics[angle=0,scale=.83]{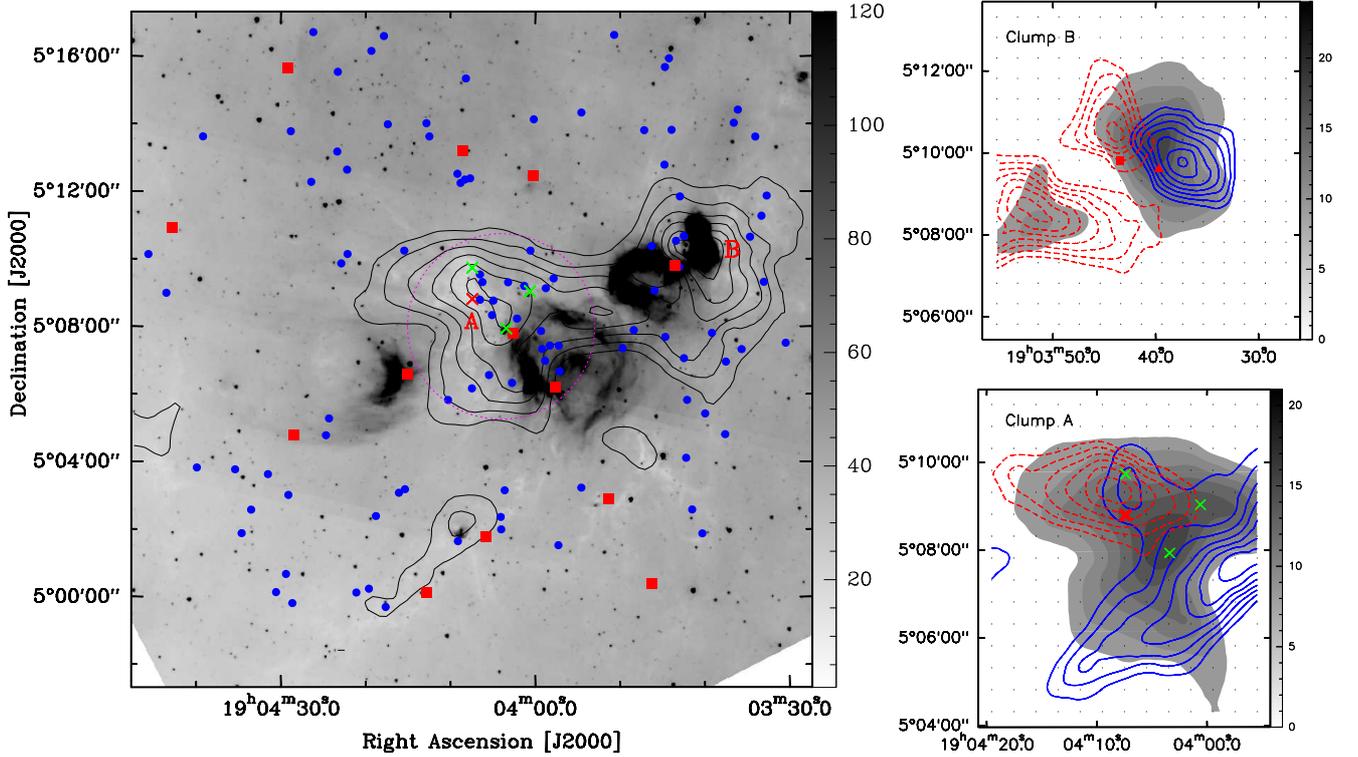}
\vspace{-27mm}\caption{Left: Class I sources are labeled as the
blue dots. The red
squares and triangle represent IRAS sources and Class II methanol maser (Deharveng et al. \cite{Deharveng10}), respectively. The green and red crosses indicate five smm continuum sources.  The purple dashed circle outlines the Class I sources, which may be associated with IRDC G38.95-0.47. Right: the velocity-integrated intensity maps of
$^{13}$CO $J=1-0$ outflows (red and blue contours) overlaid with the
$^{13}$CO $J=1-0$ emission of each clump (gray scale). The red and
blue contour levels are 30,...,100\% of the peak value. }
\end{figure*}

\begin{table*}
\begin{center}
\small \tabcolsep 2.0mm\caption{Selected IR point sources associated with these
H {\small II} regions: IR flux densities and the obtained parameters}
\begin{tabular}{clccccccccccccr}
\hline\hline Name & Source & RA & DEC & $F_{12}$ & $F_{25}$
& $F_{60}$ & $F_{100}$ &$\log(\frac{F_{25}}{F_{12}})$&$\log(\frac{F_{60}}{F_{12}})$ & $L_{\rm IR}$& $T_{\rm d}$   \\
 & &(h m s)&($\circ$ $\prime$  $\prime\prime$)& [Jy] & [Jy]& [Jy]& [Jy]& & &[$L_{\odot}$]&[K]\\
\hline
1 & IRAS 19012+0505 &19 03 43.51  &05 09 48.91 &8.27 &17.31 &497.20  &1156.00 &0.32      &1.78  &11644.14 & 91.62&\\
\hline
\end{tabular}
\end{center}
\end{table*}

\subsection{Search for young stellar objects}
To look for star formation toward IRDC G38.95-0.47 and its adjacent H {\footnotesize II} regions, we used the Spitzer-GLIMPSE I catalog. The GLIMPSE I survey observed the Galactic plane (65$^{\circ}$$<$ $|l|$ $<$ 10$^{\circ}$ for $|b|$ $<$ 1$^{\circ}$) with the four
mid-IR bands (3.6, 4.5, 5.8, and 8.0 $\mu$m) of the Infrared Array
Camera (Fazio et al. \cite{Fazio}). From the database, we extracted  9046 near-infrared
sources within a circle of 10$^{\prime}$ in radius centered on
R.A.=$19^{\rm h}04^{\rm m}07.5^{\rm s}$ (J2000),
Dec=$05^{\circ}08'18.9^{\prime\prime}$ (J2000).  Based on the color selection criteria of YSOs (Allen et al. \cite{Allen};  Robitaille et al. \cite{Robitaille}), we only selected the 103 Class I sources that had been detected in the four Spitzer-IRAC bands. Class I sources are protostars
with circumstellar envelopes that are  expected to be YSOs with an age of $\sim$10$^{5}$ yr. Figure 4 presents the spatial distribution of Class I sources. From Fig. 4, we see that Class I sources are asymmetrically distributed across the whole selected
region, and are mostly concentrated in IRDC G38.95-0.47 between
H {\footnotesize II} regions G38.91-0.44 and G39.30-1.04.  The existence of Class I sources may
indicate star formation activity.

In addition, we found 13 $IRAS$ sources using the $IRAS$ Point Sources Catalog in this region. In Fig. 4, five $IRAS$ sources may be associated with PAH emission.
Because the fluxes of the five $IRAS$ sources are only upper limits except for IRAS 19012+0505, we only calculate the parameters of IRAS 19012+0505.
Infrared luminosity (Casoli et al. \cite{Casoli86}) and
dust temperature (Henning et al. \cite{Henning90}) are expressed respectively
as,
\begin{equation}
\mathit{L}_{\rm IR}=(6.196\times F_{12}+2.261\times
F_{25}+1.373\times F_{60}+0.529\times F_{100})\times
D^{2},
\end{equation}
\begin{equation}
\mathit{T}_{\rm
d}=\frac{96}{(3+\beta)\ln(100/60)-\ln(F_{60}/F_{100})},
\end{equation}
where $D$ is the distance from the sun in kpc, and $F_{12}$, $F_{25}$, $F_{60}$, and $F_{100}$ are the infrared fluxes at
four IRAS bands, 12 $\mu$m, 25 $\mu$m, 60 $\mu$m, and 100 $\mu$m, respectively. The emissivity index
of dust particle ($\beta$) is assumed to be 2. In addition, we calculate the color index of each  $IRAS$ source.  The calculated
results are presented in  Table 2.

\section{DISCUSSION}

The $^{13}$CO $J$=1-0 emission of IRDC G38.95-0.47 shows a triangle-like shape (see Figs. 1 and 4), and has a steep integrated-intensity gradient toward H {\footnotesize II} regions G38.91-0.44 and G39.30-1.04. We suggest that H {\footnotesize II} regions G38.91-0.44 and G39.30-1.04 have expanded into the IRDC G38.95-0.47. Four submillimeter continuum sources have been detected in the IRDC G38.95-0.47. Only source G038.95-00.47-M1  has outflow and infall motions, as well as a solar mass of 117 $M_{ \odot}$, indicating indicating that it is a newly forming massive star. Both kinetic timescales of the outflow and infall are 4.7$\times$$10^{4}$ yr (L\'{o}pez-Sepulcre et al.  \cite{Lopez-Sepulcre10}).
Clump B shows a bow-like shape toward G38.93-0.39. We conclude that G38.93-0.39 has also expanded into clump B. Additionally, clump B is associated with a Class II methanol maser at 6.7 GHz (Deharveng et al. \cite{Deharveng10}). Class II methanol masers are considered  signposts of recent massive star formation, but Zhang et al. (\cite{Zhang}) did not detect the outflow in  clump B coincident with IRAS 19012+0505.  According to the color standard of Hughes \& Macleod (\cite{Hughes89}), we find that IRAS 19012+0505 is associated with H {\footnotesize II} region G38.93-0.39. To confirm and explore the outflow motions in the two clumps, we show the distribution of the integrated intensity over the wings of the $^{13}$CO $J=1-0$ profiles in Fig. 3 (right).  We identify the wings with the velocity-position diagrams. In  Fig. 3, we see that the $^{12}$CO $J=1-0$ profiles are contaminated by the emission of the background source. The velocity
component of blueshifted emission is from 39.7 km s$^{-1}$ to 40.7 km s$^{-1}$,
while the velocity component of redshifted emission is from 42.6 km s$^{-1}$ to
43.6 km s$^{-1}$ for G038.95-00.47-M1. In IRAS 19012+0505, the velocity
component of blueshifted emission is from 36.5 km s$^{-1}$ to 38.2 km s$^{-1}$,
while the velocity component of redshifted emission is from 40.2 km s$^{-1}$ to
41.0 km s$^{-1}$. In Fig. 4,
the blueshifted and redshifted components are presented as blue and
red contours.  From Fig. 4 (right), we clearly see that  clump B has a collimated outflow activity at the NE-SW direction. Moreover, there is also outflow motion in  clump A consistent with G038.95-00.47-M1. However, the blueshifted lobe of  clump  A is weak, which may be disrupted by the expansion of the  H {\footnotesize II} region G38.91-0.44.
Adopting the angle of 90$^{\circ}$, the dynamic timescale of each outflow is estimated by equation t = 9.78$\times$ $10^{5}$R/V (yr)(Goldsmith et al. \cite{Goldsmith}; Qin et al. \cite{Qin08}), where V in km $\rm s^{-1}$ is the maximum flow velocity relative to the cloud systemic velocity, and R in pc is the outflow size defined
by the length of the begin-to-end flow extension for each blueshifted and redshifted lobe. Because IRDC G38.95-0.47 is associated with H {\footnotesize II} regions G38.91-0.44 and G39.30-1.04, we also adopt a kinematic distance of 2.9 kpc in this paper. The average dynamical timescale of the outflow in clump A is  2.8$\times$$10^{5}$ yr, which is
roughly consistent with the value in HCO$^{+}$ $J$=1-0 line (L\'{o}pez-Sepulcre et al.  \cite{Lopez-Sepulcre10}). For  clump B, the outflow has the average dynamical timescale of 3.0$\times$$10^{5}$ yr.

In addition, Class I sources are mostly concentrated in IRDC G38.95-0.47 between
H {\footnotesize II} regions G38.91-0.44 and G39.30-1.04.  Along the northeastern and eastern  sides of N74 associated with G38.91-0.44, both Beaumont \& Williams (\cite{Beaumont11}) and  Alexander et al. (\cite{Alexander}) found a significant statistical overdensity of YSOs above the surrounding field. We do not know if all the selected sources seen in the direction of this region lie at the same distance as IRDC G38.95-0.47 relative to G38.91-0.44 and G39.30-1.04. However, the high density of YSOs located in the IRDC indicates that
it is unlikely that they  are all merely
foreground and background stars. It is more likely that these YSOs
are physically associated with IRDC G38.95-0.47, between
H {\footnotesize II} regions G38.91-0.44 and G39.30-1.04.

The above analysis suggest that the triggered star formation have occurred in this region. The dynamical age of the H {\small II} region can also be used to  decide whether YSOs are triggered by H {\small II} regions. Assuming an H {\small II} region expanding in a homogeneous medium, the dynamical age of the H {\small II} region can be estimated by the Dyson \& Williams (\cite{Dyson80}) equation
\begin{equation}
\mathit{t_{\rm  HII}}=7.2\times10^{4}(\frac{R_{\rm H{\small II}}}{\rm pc})^{4/3}(\frac{Q_{\rm Ly}}{10^{49} \rm ph~s^{-1}})^{-1/4}(\frac{n_{\rm i}}{10^{3}\rm cm^{-3}})^{-1/2} \rm ~yr,
\end{equation}\indent
where $R_{\rm HII}$ is the radius of the H {\small II} region, $n_{\rm i}$ is the initial number density of the gas, and $Q_{\rm Ly}$ is the ionizing luminosity.  In previous studies toward several H {\small II} regions (e.g., Zavagno et al. \cite{Zavagno06}; Deharveng et al. \cite{Deharveng08}; Paron et al. \cite{Paron09,Paron11}; Pomar\`{e}s et al. \cite{Pomares09}; Dirienzo et al. \cite{Dirienzo12}), an initial number density of $\sim$10$^{3}$cm$^{-3}$ was determined. Alexander et al. (\cite{Alexander}) gave the ionizing luminosity of 1.2$\times10^{46}$ ph s$^{-1}$, 2.1$\times10^{46}$ ph s$^{-1}$, and 6.6$\times10^{45}$ ph s$^{-1}$ for G38.91-0.44, G38.93-0.39 and  G39.30-1.04, respectively. Adopting the measured radius of H {\small II} regions G38.91-0.44 ($\sim$1.4 pc), G38.93-0.39 ($\sim$0.8 pc), and G39.30-1.04 ($\sim$1.4 pc) obtained from Fig. 1, and assuming an initial number density of $\sim$10$^{3}$cm$^{-3}$, we derived that the ages of these H {\small II} regions are 6.1$\times10^{5}$ yr, 2.5$\times10^{5}$ yr, and 9.0$\times10^{5}$ yr, respectively. Comparing the ages of these H {\small II} regions with YSOs (Class I sources and massive G038.95-00.47-M1 source), we suggest that the YSOs  located in IRDC G38.95-0.47 are likely to be triggered by  G38.91-0.44 and G39.30-1.04 together.  Deharveng et al. (\cite{Deharveng03}) suggest that some dense fragments are
regularly spaced along the H {\small II} region, providing strong evidence in favor
of the CC model. In this picture, the shock fronts of G38.91-0.44 and G39.30-1.04 have driven into clump A, and have compressed some pre-existing cores in the IRDC G38.95-0.47. Furthermore, the PAH emission of G39.30-1.04 shows the cometary
globule, supporting the RDI model. It may be the first time that the triggered massive and clustered stars formation has occurred in the IRDC compressed by two H {\small II} regions.  We detected an outflow in clump B. Because the age of G38.93-0.39 is slightly smaller that of the outflow, we conclude that the triggered star formation has not occurred in clump B, as we also did not find a significant statistical overdensity of YSOs (10$^{5}$ yr) surrounding G38.93-0.39.
Clump B is consistent with a star cluster  (Mercer et al. \cite{Mercer}), hence the outflow may be driven by the cluster.

\section{Conclusions}

We have shown the $^{12}$CO $J=2-1 $, $^{12}$CO $J=3-2$, and
$^{13}$CO $J=2-1$ molecular, infrared, and radio continuum observations towards IRDC G38.95-0.47 and its adjacent H {\footnotesize II} regions G38.91-0.44 (N74), G38.93-0.39 (N75), and G39.30-1.04. The results can be summarized as follows:
 \begin{enumerate}
      \item
 The $^{13}$CO $J$=1-0 emission shows two large cloud clumps.  The clump  associated with IRDC G38.95-0.47 shows a triangle-like shape, and has a steep integrated-intensity gradient toward H {\footnotesize II} regions G38.91-0.44 and G39.30-1.04, indicating that the two H {\footnotesize II} regions have expanded into the IRDC.
      \item
Four  submillimeter continuum sources have been detected in the IRDC G38.95-0.47. Only the G038.95-00.47-M1 source has outflow and infall motions. In addition, the selected young stellar objects (YSOs) (Class I sources) are concentrated in the IRDC G38.95-0.47, which appear to be sites of ongoing star formation. The obtained ages of the three H {\small II} regions are 6.1$\times10^{5}$ yr, 2.5$\times10^{5}$ yr, and 9.0$\times10^{5}$ yr.  Taking into account the age of H {\footnotesize II} regions and YSOs (Class I sources and massive G038.95-00.47-M1 source), we suggest that YSOs may be triggered by the combined energy G38.91-0.44 and G39.30-1.04, supporting the radiatively driven implosion model. It may be the first time that the triggered star formation has occurred in the IRDC compressed by two H {\small II} regions.
      \item
We detected a new collimated outflow in the clump compressed by G38.93-0.39. The new detected outflow may be driven by a star cluster.

   \end{enumerate}

\begin{acknowledgements} We are very grateful to the anonymous referee for his/her helpful comments and suggestions. We are also grateful to the staff at the Qinghai Station of PMO for
their assistance during the observations. Thanks for the Key Laboratory
for Radio Astronomy, CAS, for partly supporting the telescope
operation. This work was supported by the National Natural Science Foundation of China (Grant No. 11363004).
\end{acknowledgements}

\end{document}